\documentclass
[preprint,prb,a4paper,amsfonts,amssymb,floats,titlepage,fleqn,footinbib,lengthcheck,twocolumn,tightenlines,balancelastpage,10pt,superscriptaddress,showpacs]{revtex4-1}%
\usepackage{natbib}
\usepackage{amssymb}
\usepackage{amsfonts}
\usepackage{amsmath}
\usepackage{graphicx}
\usepackage{color}
\usepackage{longtable}
\usepackage{revsymb}
\usepackage{units}
\usepackage{nicefrac}
\usepackage{textcomp}
\usepackage{epsfig}
\usepackage{footmisc}
\usepackage{soul}%
\ifx\pdfoutput\relax\let\pdfoutput=\undefined\fi
\newcount\msipdfoutput
\ifx\pdfoutput\undefined\else
\ifcase\pdfoutput\else
\ifx\paperwidth\undefined\else
\ifdim\paperheight=0pt\relax\else\pdfpageheight\paperheight\fi
\ifdim\paperwidth=0pt\relax\else\pdfpagewidth\paperwidth\fi
\fi\fi\fi
\begin{document}
\title{Pressure-dependent magnetization and magnetoresistivity studies on the
tetragonal FeS (mackinawite): revealing its intrinsic metallic character}
\author{S. J. Denholme}
\affiliation{National Institute for Materials Science, 1-2-1, Sengen, Tsukuba, 305-0047, Japan}
\author{H. Okazaki}
\affiliation{National Institute for Materials Science, 1-2-1, Sengen, Tsukuba, 305-0047, Japan}
\author{S. Demura}
\affiliation{National Institute for Materials Science, 1-2-1, Sengen, Tsukuba, 305-0047, Japan}
\author{K. Deguchi}
\affiliation{National Institute for Materials Science, 1-2-1, Sengen, Tsukuba, 305-0047, Japan}
\author{M. Fujioka}
\affiliation{National Institute for Materials Science, 1-2-1, Sengen, Tsukuba, 305-0047, Japan}
\author{T. Yamaguchi}
\affiliation{National Institute for Materials Science, 1-2-1, Sengen, Tsukuba, 305-0047, Japan}
\author{H. Takeya }
\affiliation{National Institute for Materials Science, 1-2-1, Sengen, Tsukuba, 305-0047, Japan}
\author{M. ElMassalami}
\affiliation{Instituto de F\'{\i}sica, Universidade Federal do Rio de Janeiro, Caixa Postal
68528, 21941-972 Rio de Janeiro RJ, Brazil}
\author{H. Fujiwara}
\affiliation{Research laboratory for surface science, Okayama University, Okayama,}
\author{T. Wakita}
\affiliation{Research laboratory for surface science, Okayama University, Okayama,
700-8530, Japan}
\author{T. Yokoya}
\affiliation{Research laboratory for surface science, Okayama University, Okayama,
700-8530, Japan}
\author{Y. Takano}
\affiliation{National Institute for Materials Science, 1-2-1, Sengen, Tsukuba, 305-0047, Japan}

\begin{abstract}
The transport and magnetic properties of the tetragonal Fe$_{1+\delta}$S were
investigated using magnetoresistivity and magnetization within 2$\leq T\leq
$300 K, $H\leq$70 kOe and $P\leq$ 3.0 GPa. In addition, room-temperature X-ray
diffraction and photoelectron spectroscopy were also applied. In contrast to
previously reported nonmetallic character, Fe$_{1+\delta}$S is intrinsically
metallic but due to a presence of a weak localization such metallic character
is not exhibited below room temperature. An applied pressure reduces strongly
this additional resistive contribution and as such enhances the temperature
range of the metallic character which, for $\sim$3 GPa, is evident down to 75
K. The absence of superconductivity as well as the mechanism behind the weak
localization will be discussed.

\end{abstract}

\maketitle

\section{Introduction}

Although the isomorphous Fe-based chalcogenides Fe$_{1+\delta}X$ ($X$= Te, Se,
S) crystallize into the room-temperature tetragonal $P4/nmm$ structure,
however (as far as the \textit{low-temperature} structural, magnetic, and
electronic properties are concerned) their phase diagrams
\cite{Mizuguchi10-FeT-Review,Deguchi12-FeX-Review,Subedi08-FeX,Fang08-Fe(SeTe),Liu10-Fe(TeSe)-PhaseDiagram,Dong11-PhaseDiagram-FeTeSe,Kawasaki12-Fe(TeSe)-O-anneal-PhaseDiagram}
are distinctly different: Fe$_{1+\delta}$Se is an orthorhombic nonmagnetic
superconductor; Fe$_{1+\delta}$Te is a monoclinic antiferromagnetic metal
while Fe$_{1+\delta}$S is a tetragonal nonmagnetic and nonconducting
compound\cite{Bertaut65,Goodenough78-Chemistry-FeS,Wilson10-Prespective-Fe-SUC}
though, in sharp contrast, most theoretical work suggests metallic
character.\cite{Kwon11,Subedi08,Devy08} Further distinction among these
Fe$_{1+\delta}X$ is evident in the response of their individual states to
applied pressure, doping, intercalation, or a magnetic field. Most striking
are the differences among the superconducting phase diagrams of their solid
solutions, e.g. Fe$_{1+\delta}$(Te$_{1-x}X_{x}$) ($X$= Se,
S):\cite{Mizuguchi10-FeT-Review,Deguchi12-FeX-Review} substitution leads to a
gradual suppression of magnetism and to an eventual surge of
superconductivity; on the other hand, for Fe$_{1+\delta}$(Se$_{1-x}$S$_{x}$),
substitution leads to a slight enhancement in $T_{c}$ up to $x$=0.2 but on
further substitution the superconducting transition is monotonically suppressed.

It is remarkable that in spite of such a distinction between the electronic
states of these Fe$_{1+\delta}X$ compounds, theoretical
studies\cite{Kwon11,Subedi08,Devy08} predicted\ a metallic normal-state: while
this metallicity is established for \textit{low-temperature} phases of
Fe$_{1+\delta}$Te and Fe$_{1+\delta}$Se, experimentally Fe$_{1+\delta}$S was
reported to manifest an absence of metallic conductivity.\cite{Bertaut65} As
that the question of the electronic character of Fe$_{1+\delta}$S is of a
fundamental importance to the general understanding of the normal and
superconducting phase diagrams of these Fe-based chalcogenides, this work
addresses the electronic properties of Fe$_{1+\delta}$S using X-ray
diffraction, spectroscopic (ultra violet photoelectron spectroscopy, UPS), and
thermodynamic (magnetoresistivity and magnetization over a wide range of
temperature $T$, pressure $P$, and magnetic field $H$) techniques

Based on the stoichiometry of the iron monosulfides FeS, there are, in
general, three classes:\cite{Goodenough78-Chemistry-FeS} (i) this
Fe$_{1+\delta}$S system (mackinawite) which, just as for the other isomorphous
Fe$_{1+\delta}X$, manifests an excess of Fe and crystallizes in the layered
anti-PbO type structure;\cite{Kjekshus72-FeS} an application of 3.3 GPa at
room temperature transforms its tetragonal phase into an orthorhombic
structure.\cite{Ehm09} (ii) the near-stoichiometric and hexagonal
antiferromagnetic FeS (troilite),\cite{Horwood76,Gosselin76} and (iii) the
hexagonal Fe--deficient ferromagnetic Fe$_{1-\delta}$S ($x\leq$0.2) which
crystallizes in the nickel arsenide form (pyrrohotite).\cite{Waldner05} The
structural and physical properties of both FeS and Fe$_{1-\delta}$S have been
extensively investigated;\cite{Waldner05,Kobayashi05} in contrast,
Fe$_{1+\delta}$S has been relatively unexplored except for some structural and
mineralogical studies:\cite{Berner62,Lennie95,Vaughan71} neutron diffraction
and M\"{o}ssbauer analysis indicated nonmagnetic character;\cite{Bertaut65}
this contradicts an analysis done by photoelectron spectroscopy
(PES)\cite{Devy08} which suggested, instead, a single-stripe antiferromagnetic
ground state.

Electronic structure calculations\cite{Welz87,Kwon11,Subedi08,Devy08} on
Fe$_{1+\delta}$S indicated a significant Fe 3$d$ orbital delocalization
(primarily due to the basal-plane, intralayer Fe-Fe interactions), a dominant
3$d$ contribution to the density of states (DOS, $N(E_{F})$) at the Fermi
level, $E_{F}$, and a weaker hybridization between the Fe and
S.\cite{Kwon11,Subedi08,Devy08} As mentioned above such a predicted metallic
character is in disagreement with the experimentally observed nonmetallicity.
In this work, we show that Fe$_{1+\delta}$S is indeed metallic just as
theoretically predicted; the reported
nonmetallicity\cite{Bertaut65,Denholme14-FeS-semiconductivity} will be shown
to be due to a localization of charge carriers. It is recalled that such a
discrepancy between experiment and theory had already been reported in other
transition metal sulfides:\cite{Rohrbach03} e.g., troilite is a p-type
semiconductor with a band gap of 0.04 eV;\cite{Gosselin76} yet band structure
calculations have placed $E_{F}$ within the $d$-$p$ hybridized
bands.\cite{Ikeda93} Similarly, a PES study on pyrrohotite reported a 25-30\%
narrower Fe $3d$ DOS\ band-width than the theoretical
prediction.\cite{Shimada98}

\section{Experimental}

Mackinawite was synthesized using the method reported by Lennie \textit{et
al}.\cite{Lennie95} Powder X-ray diffractograms on a conventional Cu
\textit{K}$\alpha$ diffractometer indicated a single phase $P4/nmm$ structure
with $a$=3.675(2) \AA , $c$=5.035(6) \AA . Based on an energy dispersive X-ray
analysis, the actual stoichiometry was found to be Fe:S=0.52:0.48 giving
Fe$_{1.08}$S which is in agreement with the reported
ranges.\cite{Lennie95,Taylor70-Fe-Stoichiometry}

It is well-known that the tetragonal Fe$_{1+\delta}$S is chemically unstable
against a variation in $P$, $T$ and aging:\cite{Csakberenyi-Malasics12} aging
at room temperature would slowly transform it into an amorphous product plus
the semi-metallic cubic Fe$_{3}$S$_{4}$\ (greigite). During this study, it
became evident that (i) such a conversion can be temporarily inhibited if the
sample is stored at cooler temperatures e.g. below 5 C$^{\circ}$; (ii) this
tetragonal Fe$_{1+\delta}$S, when subjected to a higher pressure, would start
to convert into an amorphous product, reminiscent of the amorphization in
Fe$_{1+\delta}$Se and FeSe$_{0.5}$Te$_{0.5}$%
,\cite{Stemshorn10-FeSe-HighPressure,*Stemshorn09-FeSe0.5Te0.5.-highPressure}
plus the semiconducting hexagonal Fe$_{1-\delta}$S (troilite). Accordingly,
such a phase instability requires that extra care should be exercised during
(as well as before and after) the measurements so as to ensure that all
results had been obtained on the very same tetragonal phase: otherwise most of
the results (in particular the resistivity) are irreproducible. With this in
mind, the following measurements and their analysis were carried out.

Resistivity, $\rho$, was measured using a standard four-in-line method on
cold-pressed pellets (care was undertaken to ensure that the grain boundary
influence was minimized - see below). For $\rho(P)$, hydrostatic pressures, up
to 3.0 GPa, were generated by a BeCu/NiCrAl clamped piston-cylinder cell using
Fluorinert as a $P$-transmitting fluid while Pb as a manometer. Similarly,
$P$-dependent magnetizations were measured using a hydrostatic pressure cell
(up to 1 GPa). Daphne oil was used as a $P$-transmitting fluid while Sn as a manometer.

UPS was measured at a base pressure of 2.0$\times$10$^{-8}$ Pa and at a
temperature of 300K with He I (21.2 eV), He II (40.8eV) and Xe I (8.44 eV)
resonance lines. So as to obtain a fresh surface, samples were cut within an
ultra-high vacuum chamber. The Fermi energy was referenced to that of an Au
film which was measured frequently during the experiments.

\section{Results}%

\begin{figure}[tb]%
\centering
\includegraphics[
height=5.9089cm,
width=7.7079cm
]%
{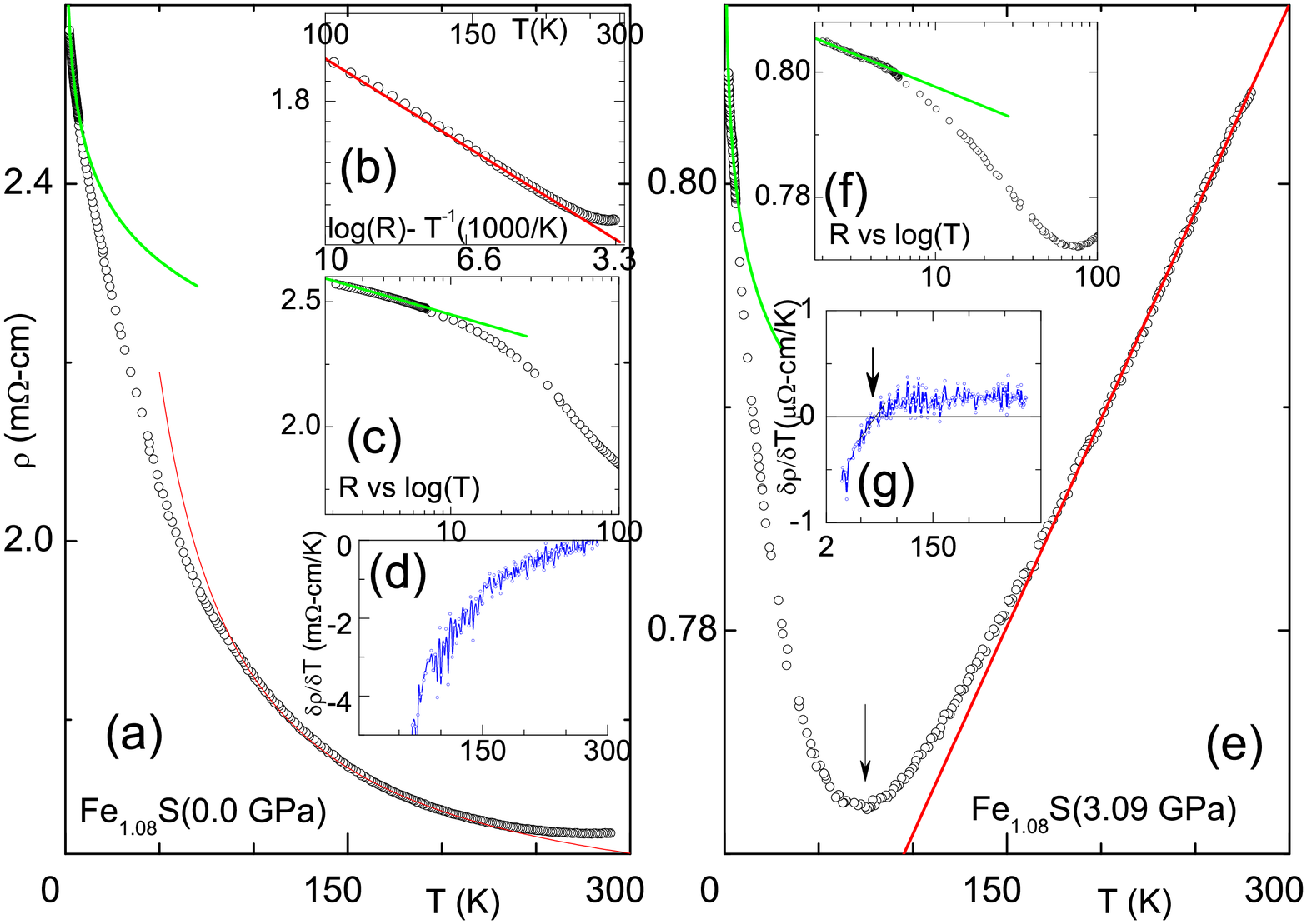}%
\caption{Typical resistivity curves of Fe$_{1.08}$S (a-d) under
ambient-pressure while (e-g) are under an applied pressure of 3.09 GPa. (a and
e) isobaric $\rho$ $vs$ $T$ curves. (b) $\rho$ $vs$ $T$ curve in a
log-reciprocal plot emphasizing the activated process; the solid line is a fit
to Eq.\ref{Eq.Arrhenuis}. (c and f) $\rho$ $vs$ $T$ curves in a linear-log
plot emphasizing the weak localization behavior; the solid lines is a fit to
Eq. \ref{Eq.Localization}. (d and g) $\frac{\partial\rho}{\partial T}$ $vs$
$T$ curve: a negative (positive) value represents nonmetallic- (metallic-)
like behavior. The crossover point is denoted as $T_{L}$. The obtained $T_{L}%
$, $\frac{\partial\rho}{\partial T},$ $\Delta$, $S$ parameters are collected
in Fig. \ref{Fig-FeS-pt-gap-s-phasediagram}.}%
\label{Fig-FeS-RvT-pressure}%
\end{figure}
Fig. \ref{Fig-FeS-RvT-pressure}(a) illustrates $\rho(T,H$=0, $P$=0.1 MPa$)$ of
Fe$_{1.08}$S.\cite{Bertaut65,Denholme14-FeS-semiconductivity} It is remarkable
that $\rho($2$\leq T\leq$300 K$)${ }${\sim}${m}$\Omega$-cm{ suggesting that
this (monotonic but non-sharp) low-}$T${ rise is not due to a conventional
metal-insulator transition; most probably, it is a manifestation of
localization of charge
carriers\cite{Anderson79-NonlinearConductance,*Dolan79-nonlinearConductance,Fukuyama80-Interaction-2D-localization,*Fukuyama83-Interaction-2D-localization,Altshuler83-Localization-e-e}
(see below). Following the analysis of Ref. \onlinecite{14-PhaseDiagram-FeX},
it is taken that }the{ resistivity is intrinsically metallic, any
nonmetallicity is attributed to this localization.} Fig.
\ref{Fig-FeS-RvT-pressure}(d) reveals such nonmetallic character as a negative
$\frac{\partial\rho}{\partial T}$. Moreover, as $T\rightarrow$300K,
$\frac{\partial\rho}{\partial T}$ $\rightarrow0$ at $\sim$300K: assuming a
stable tetragonal phase (see Experimental), the event $\frac{\partial\rho
}{\partial T}$=0 is taken as a crossover from a nonmetallic\ state into a
metallic one. A closer look at the evolution of $\rho(T<$300 K$)$ suggests
that there are at least two types of localization-induced
behavior:\cite{14-PhaseDiagram-FeX} The first {appears to be} a
thermally-assisted behavior:\cite{Mott68-MIT,*Mott87-MobilityEdge}
\begin{equation}
\rho(\text{100}<T<\text{300K})=\rho_{0}^{A}\exp(\bigtriangleup
/T).\label{Eq.Arrhenuis}%
\end{equation}
Such a thermally activated ($\bigtriangleup$ $\sim$ 20 K) process [see Fig.
\ref{Fig-FeS-RvT-pressure}(b)] is assumed to be due to a hopping of carriers
from one localized state into an itinerant state that is separated by an
effective $\bigtriangleup=\left\vert E-E_{c}\right\vert $ where $E_{c}$ is the
mobility edge\cite{Mott68-MIT,*Mott87-MobilityEdge} and should not be confused
with the conventional semiconducting behavior. A manifestation of an activated
behavior below a crossover/transition was already observed in \textrm{RNiO}%
$_{3}$.\cite{Obradors93-MIT-RNiO3,*Granados93-MIT-NdNiO3} The second process{
looks like a} weak-localization process which is{ often encountered} in the
low-$T$ phase of a disordered
metal.\cite{Anderson79-NonlinearConductance,*Dolan79-nonlinearConductance,Fukuyama80-Interaction-2D-localization,*Fukuyama83-Interaction-2D-localization,Altshuler83-Localization-e-e}
In chalcogenides,\cite{14-PhaseDiagram-FeX} the disorder is attributed to the
nonperiodic scattering potentials (see below). Then their $\rho(T)$ should
follow\cite{Anderson79-NonlinearConductance,*Dolan79-nonlinearConductance,Fukuyama80-Interaction-2D-localization,*Fukuyama83-Interaction-2D-localization,Altshuler83-Localization-e-e}
\begin{equation}
\rho(T<20K)=\rho_{o}^{L}[1+S\ln(T_{o}/T)]\label{Eq.Localization}%
\end{equation}
where $S$ is a measure of the scattering process while $T_{o}$ and $\rho
_{o}^{L}$ are any measured pair [Fig. \ref{Fig-FeS-RvT-pressure}(c)]. Such a
log-in-$T$ character was already reported for other
chalcogenides.\cite{14-PhaseDiagram-FeX,Liu09-FeExcess-FeTeSe,*Liu10-Fe(TeSe)-PhaseDiagram,*Chang12-FeTe1-xSex-WeakLocalization}
Presently it is not evident why this process is not proceeded by a
metallic-like state as observed in, e.g.,
Ref.[{\onlinecite{14-PhaseDiagram-FeX}}].

An application of pressure leads to pronounced effects: e.g. (i) on comparing
$\rho(T)$ of panels (a) and (e) of Fig. \ref{Fig-FeS-RvT-pressure}, one
notices a reduction in the overall resistivity, and (ii) the crossover point
signalled by $\frac{\partial\rho}{\partial T}$=0, denoted as $T_{L}(P)$, moves
to well below 300K: the metallicity is pressure-enhanced to a wide range of
temperature.\cite{Obradors93-MIT-RNiO3,*Granados93-MIT-NdNiO3}

\begin{figure}[tb]%
\centering
\includegraphics[
height=5.6585cm,
width=7.7079cm
]%
{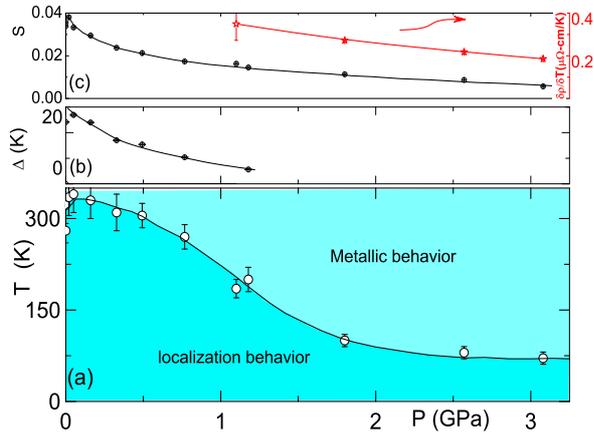}%
\caption{The baric evolution of (a) $T_{L}$ of Fe$_{1.08}$S\ (a measure of
localization strength which is emphasized below $T_{L}(P)$ curve), (b)
$\Delta$ (the activation energy in Eq. \ref{Eq.Arrhenuis}) (c) \textit{left
ordinate:} $S$ (as a measure of strength of the localization process below 20
K, see Eq. \ref{Eq.Localization}); \textit{right ordinate:} $\frac
{\partial\rho}{\partial T}$ within $T_{L}<T<$ 300 K (a measure of the thermal
evolution of the metallic resistivity).}%
\label{Fig-FeS-pt-gap-s-phasediagram}%
\end{figure}

Just as for the ambient-pressure case, $\rho(T<T_{L},P)$ of Fig.
\ref{Fig-FeS-RvT-pressure}(e) was analyzed in terms of the above mentioned two
processes. The baric evolution of the fit parameters are shown in Fig.
\ref{Fig-FeS-pt-gap-s-phasediagram}; $P$ reduces all scattering processes: a
monotonic decrease of (i) $\frac{\partial\rho}{\partial T}$ within the
metallic state, (ii) $\Delta$ within the activated region, and (iii) $S$ below
20 K. All these influences lead to a strong reduction of $T_{L}(P<$2GPa$)$.
Above 2 GPa, $T_{L}(P)$ is weakly but monotonically decreasing till $T_{L}%
(P=$3.1GPa$)\sim$75 K; such a thermal evolution is also manifested for each of
the parameters shown in Fig. \ref{Fig-FeS-pt-gap-s-phasediagram}(b-c).
\begin{figure}[tb]%
\centering
\includegraphics[
height=5.307cm,
width=7.7079cm
]%
{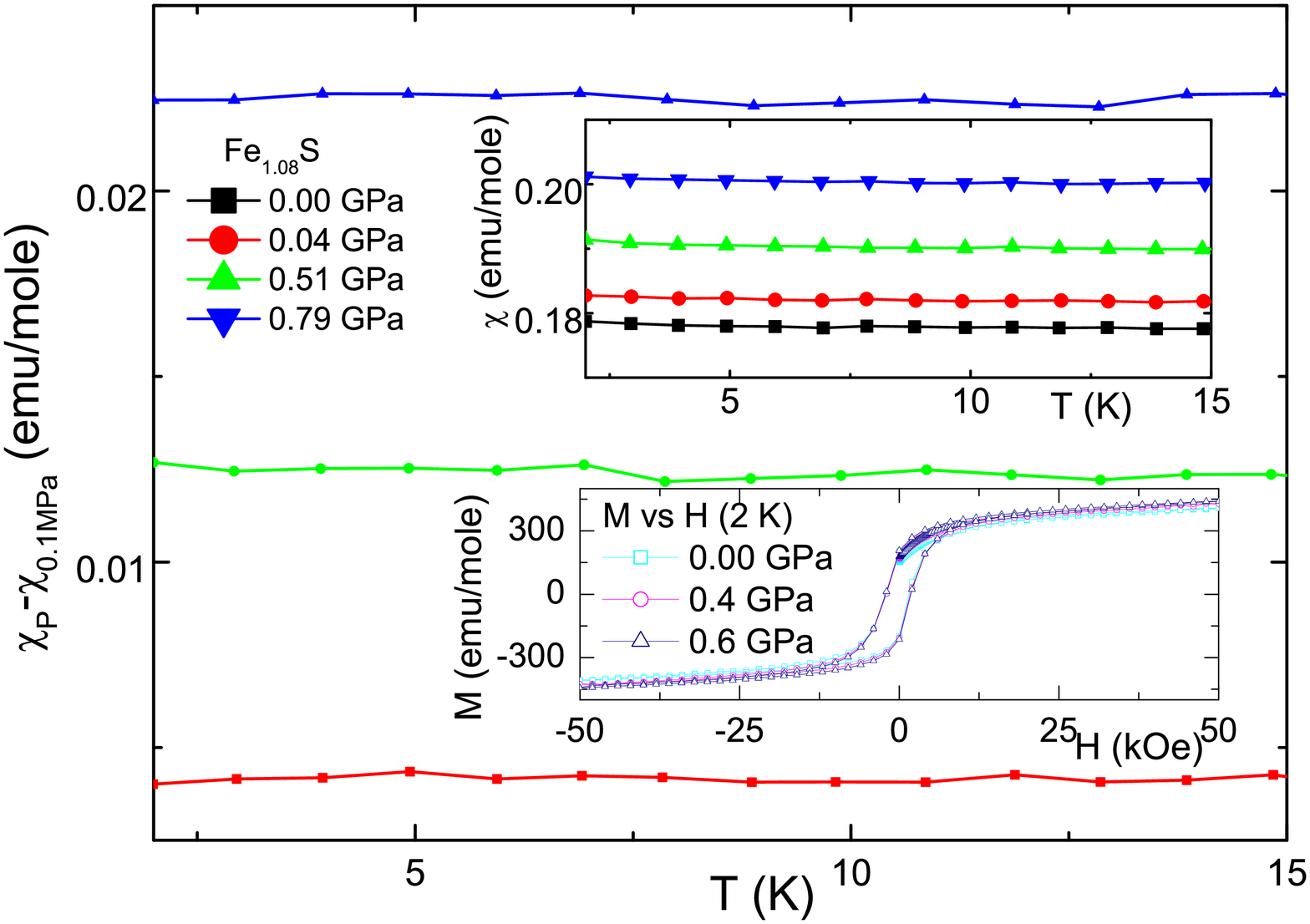}%
\caption{The excess, pressure-dependent molar susceptibility of Fe$_{1.08}$S.
The ambient\ pressure $\chi(T$,P=0.0 GPa$)$ curve (upper inset) is similar to
that of Sines \textit{et al.}\cite{Sines12-tetragonal-FeS} \ As the total
contribution includes those of weak magnetic impurities and the cell body and
as that these contributions are not influenced by $P${ (see the two insets)},
then, for clarity, these contributions are subtracted out by plotting, in the
main frame, $\bigtriangleup\chi(T,P)=\chi(T,P)-\chi(T,$0.0 GPa$)$.}%
\label{Fig-FeS-mvst-pressure}%
\end{figure}

Fig.\ref{Fig-FeS-pt-gap-s-phasediagram}(a) identifies unambiguously the
metallic state as being an intrinsic high-$T$ property of Fe$_{1+\delta}$S:
this provides a direct confirmation of the theoretical predictions. Evidently,
without the clarification provided by the high-$P$ or high-$T$ $\rho(T,P)$
curves, the activated rise in $\rho$($T<$300 K) as $T$ is lowered would be
mistakenly taken as indicative of intrinsic nonmetallic
conductivity.\cite{Bertaut65,Denholme14-FeS-semiconductivity,Goodenough78-Chemistry-FeS,Wilson10-Prespective-Fe-SUC}%

Various possible mechanisms can give rise to the pressure influence on each of
the $\frac{\partial\rho}{\partial T},$ $\Delta$, $S$ parameters (Fig.
\ref{Fig-FeS-pt-gap-s-phasediagram}); two of which are (i) the cold-pressed
pelletizing process brings together the already metallic grains; as such an
application of further pressure (during the $\rho(T,P)$ measurement) would
lead to a further enhancement of the grain connectivity. (ii) The influence of
the pressure is intrinsic both on the involved scattering processes as well as
on the electronic structure of Fe$_{1+\delta}$S. To differentiate between
which of these is the most plausible mechanism, we carried out a magnetization
measurements. Based on the above, Fe$_{1+\delta}$S is expected to be a
nonmagnetic metal, thus its $\chi(T)$ should be constant-in-$T$, Pauli-like
and proportional to $N(E_{F})$. If the observed $P$-induced effects are due to
grains connectivity then $\chi(T)$ should not be influenced. If, otherwise,
$P$ influences its nonpolarized electronic structure, then its $\chi(T)$
should also be modified. Indeed Fig. \ref{Fig-FeS-mvst-pressure} indicates a
$P$-induced enhancement of $\chi$ and as such an enhancement of $N(E_{F})$:
this is in an excellent agreement with the $P$-induced enhancement of the
conductivity observed in Fig. \ref{Fig-FeS-RvT-pressure}.

The general features of Figs. \ref{Fig-FeS-RvT-pressure}%
-\ref{Fig-FeS-pt-gap-s-phasediagram} can also be interpreted in terms of a
$P$-induced reduction of the involved scattering processes (not only as an
enhancement of $N(E_{F})$ as in Fig. \ref{Fig-FeS-mvst-pressure}); this
suggests that\ (i) within the metallic state ($T>T_{L}$), the electron-phonon
or elelctron-electron interactions are reduced and that (ii) for $T<T_{L}$,
$P$ induces a partial (but weak) delocalization of those carriers that had
been previously localized.

\begin{figure}[tb]%
\centering
\includegraphics[
height=4.3317cm,
width=7.8288cm
]%
{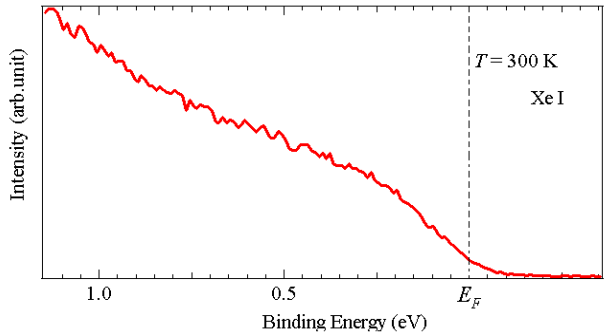}%
\caption{UPS spectrum of Fe$_{1.08}$S near $E_{F}$ (dashed vertical line)
using a Xe I source ($hv$ = 8.44 eV), under ultrahigh vacuum at $T$=300 K.}%
\label{Fig-FeS-UPS-Spectra}%
\end{figure}

From above it is concluded that\ Fe$_{1+\delta}$S is intrinsically metallic
but below $T_{L}$ localization effects are manifested. Then it is interesting
to investigate the influence of localization on the electronic states at the
Fermi surface. We addressed this question by carrying out a PES study using
UPS. Fig \ref{Fig-FeS-UPS-Spectra} shows a UPS spectra near $E_{F}$ measured
using a Xe I source of $hv$ = 8.44 eV under ultrahigh vacuum at $T=$300 K. We
observed a broad spectra with no Fermi edge: a non-metallic state which should
be contrasted with the metallic features observed in Fe$_{1+\delta}%
$Te.\cite{Xia09-Fe(TeSe)-FS-QuasiParticle} The presence of localization in
this system could account for these features. We observed the same phenomenon
with the He I (21.2 eV) and He II (40.8 eV) spectra but given the greater mean
free path of the Xe I source (ca. 1$\mu$m) we take this result as being more
representative of the bulk sample.

\begin{figure}[tb]%
\centering
\includegraphics[
height=5.5742cm,
width=7.7063cm
]%
{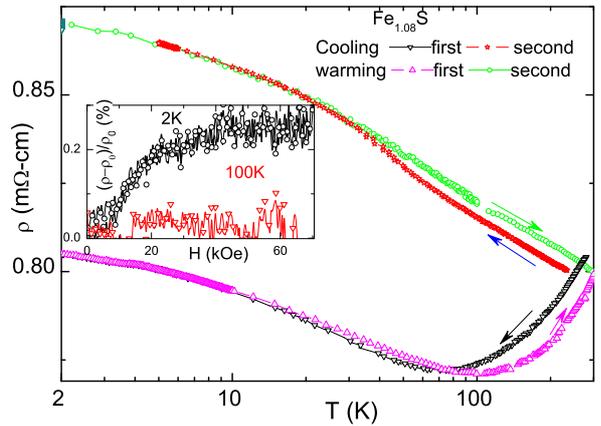}%
\caption{Various resistivity curves of Fe$_{1.08}$S measured under 3.0 GPa:
the resistivity curve\ during the first cooling is similar to the first
warming (taking into consideration the inherent thermal lag due to the massive
body of the pressure cell). On a second cooling, the resistivity deviates
strongly from the first cooling curve: this is related to phase instability.
On a final warming branch, the resistivity retraces the behavior of the second
cooling. The inset shows that the magnetoresistivity measured at 2 K (circles)
and 100 K (triangle).}%
\label{Fig-FeS-RvsH}%
\end{figure}

The phase instability of Fe$_{1+\delta}$S can be best illustrated by
$\rho(T,P)$ of Fig. \ref{Fig-FeS-RvsH}: on a first cooling branch, $\rho
(T$,3GPa$)$ shows metallic behavior followed by a localization-induced uprise
below $T_{L}$. On warming, $\rho(T,$3GPa$)$ follows the cooling curve except
at high-$T$ wherein thermal lag is manifested due to thermal gradients that
are generated across the massive body of the pressure cell. On a second
cooling branch, after some days at room temperature, $\rho(T,$3GPa$)$ was
found to be completely modified, showing an absence of metallic-like behavior
and an uprise on lowering the temperature which starts already at 300K. This
irreproducibility is related to the above mentioned phase instability: indeed
post-measurement XRD data indicated a partial phase transformation to the
hexagonal troilite form but with no evident change in the lattice parameters
of the remaining mackinawite phase. Evidence of amorphous material (most
probably amorphous mackinawite) was also found.%

\section{Discussion and conclusions}

Fig. \ref{Fig-FeS-pt-gap-s-phasediagram}(a) is a manifestation of a
pressure-induced enhancement of the stability of the metallic
phase.\cite{Okada09-PhaseTrans-FeTe} Similar enhancement had been observed in
the $R$NiO$_{3}$ charge-transfer
perovskites,\cite{Obradors93-MIT-RNiO3,*Granados93-MIT-NdNiO3} wherein a
metal-insulator transition at $T_{\text{MIT}}$ marks the sharp uprise in
$\rho(T<$ $T_{\text{MIT}})$ and, furthermore, the monotonic decrease in
$T_{\text{MIT}}(P)$ is related to the $P$-induced decrease in the
charge-transfer gap. In spite of the similarity in the manifestation of the
localization and the associated $P$-induced effects, we believe that the
crossover event in Fe$_{1+\delta}$S is not due to an Anderson-type
metal-insulator transition because (i) the is no low-$T$ AF order or a strong
hysteresis effects, (ii) the rise in $\rho(T<$ $T_{L})$ is weak (${\sim}$%
{m}$\Omega$-cm), smooth and extends over a wider temperature range, and (iii)
there are two (an activated and a log-in-$T$ ) processes operating at
different temperature regions. Instead, it is assumed that there is a weak
localization process which is due to scattering from any non-periodically
arranged potentials (the most probable disorder/defects centres are the
randomly distributed excess Fe or chalcogens deficiencies as in Fe$_{1+\delta
}$Te and Fe$_{1+\delta}$%
Se).\cite{14-PhaseDiagram-FeX,Liu09-FeExcess-FeTeSe,*Liu10-Fe(TeSe)-PhaseDiagram,*Chang12-FeTe1-xSex-WeakLocalization}%

The manifestation of two types of localization processes is not unique to
Fe$_{1+\delta}$S; it had been already observed in thin
films\cite{Dolan79-nonlinearConductance} though the order of appearance,
as$\ T$ is varied, is inverted. The activated behavior (Eq. \ref{Eq.Arrhenuis}%
) within 100 K$<T<T_{L}$ is taken to be due to a hopping of carriers. On the
other hand, the log-in-$T$ (Eq. \ref{Eq.Localization}) behavior below 20 K is
attributed to {quantum corrections arising from} scattering from the above
mentioned non-periodic
potentials.\cite{Anderson79-NonlinearConductance,Dolan79-nonlinearConductance}

The log-in-$T$
relation\cite{Anderson79-NonlinearConductance,*Dolan79-nonlinearConductance}
is also valid for localization of weakly interacting carriers though with a
different logarithmic
prefactor.\cite{Fukuyama80-Interaction-2D-localization,*Fukuyama83-Interaction-2D-localization}
Alternatively, interaction effects in disordered 2D Fermi systems within the
metallic regime can also give rise to a log-in-$T$
relation.\cite{Altshuler83-Localization-e-e} As is the usual practice, a
distinction between whether a log-in-$T$ behavior is due to either a
non-interacting\cite{Anderson79-NonlinearConductance,Fukuyama80-Interaction-2D-localization,*Fukuyama83-Interaction-2D-localization}
or an interacting carriers can be obtained from a magnetoresistivity
experiment: on increasing $H$, a negative magnetoresistivity is manifested for
the weak localization case while a positive one for the interaction case. The
magnetoresistivity of Fe$_{1.08}$S at 2 K (inset of Fig. \ref{Fig-FeS-RvsH})
is positive indicating that interactions among the diffusing carriers are
important.\cite{Altshuler83-Localization-e-e} Such a manifestation of
electron-electron interactions is taken to be behind the absence of
superconductivity in Fe$_{1+\delta}$S: indeed no such strong field-dependent
magnetoresistivity had been observed in the isomorphous Fe$_{1+\delta}$Se (see
Fig.5 of Ref. {\onlinecite{Mizuguchi08-FeSe-27K-pressure}}). At higher
temperature ($>$100 K), such a positive magnetoresistivity is drastically
reduced while, at higher field, there is a tendency towards negative magnetoresistivity.

In summary, Fe$_{1+\delta}$S is shown to be a metal but due to localization
processes, such metallicity is not reflected in the thermal evolution of
$\rho(T<$ 300K$)$ nor in the UPS spectra. Applied pressure does reduce the
influence of the localization processes and as such the metallic character is
manifested even for temperatures as low as 75 K at 3.0 GPa. Such a pressure
influence is also evident in the Pauli-like susceptibility which is enhanced
monotonically with $P$. Using low-$T$ magnetoresistivity analysis, the weak
localization that gives rise to a log-in-$T$ behavior is suggested to be due
to interaction effects in this disordered Fe-based system. It is assumed that
such electron-electron interactions are behind the absence of
superconductivity in this Fe-based chalcogenide.

\begin{acknowledgments}
This work was supported in part by the Japan Society for the Promotion of
Science (JSPS) and the Japan Science and Technology (JST) agency through the
Strategic International Collaborative Research Program (SICORP-EU Japan).
\end{acknowledgments}

\bibliographystyle{apsrev4-1}
\bibliography{FeS-mackinawite,intermetallic,InterplaySupMag,Localization,massalami,pnictides,spinfluctuation,SupClassic}

\end{document}